\documentclass[letter,twocolumn]{jpsj3}
\def\runauthor{D. Aoki \textit{et al.}}

\title{First Observation of Quantum Oscillations in the Ferromagnetic Superconductor UCoGe}

\author{%
Dai~\textsc{Aoki}$^1$\thanks{E-mail address: dai.aoki@cea.fr}, %
Ilya~\textsc{Sheikin}$^2$,
Tatsuma~D.~\textsc{Matsuda}$^{1,3}$,
Valentin~\textsc{Taufour}$^1$,
Georg~\textsc{Knebel}$^1$, and
Jacques~\textsc{Flouquet}$^1$
}

\inst{%
$^1$INAC/SPSMS, CEA-Grenoble, 17 rue des Martyrs, 38054 Grenoble, France\\
$^2$LNCMI-G, CNRS, 25 Rue des Martyrs, 38042 Grenoble, France\\
$^3$Advanced Science Research Center, Japan Atomic Energy Agency, Tokai, Ibaraki 319-1195, Japan\\
}

\abst{%
We succeeded in growing high quality single crystals of the ferromagnetic superconductor UCoGe
and measured the magnetoresistance at fields up to $34\,{\rm T}$.
The Shubnikov-de Haas signal was observed for the first time in a U-111 system (UTGe, UTSi, T: transition metal).
A small pocket Fermi surface ($F \sim 1\,{\rm kT}$) with large cyclotron effective mass $25\,m_0$ 
was detected at high fields above $22\,{\rm T}$,
implying that UCoGe is a low carrier system accompanyed with heavy quasi-particles.
The observed frequency decreases with increasing fields,
indicating that the volume of detected Fermi surface changes nonlinearly with field.
The cyclotron mass also decreases, which is consistent with the decrease of the $A$ coefficient of resistivity.
}

\kword{unconventional superconductivity, Fermi surface, Shubnikov-de Haas effect, upper critical field, quantum criticality, effective mass, ferromagnetism, UCoGe}

\begin{document}
\maketitle
The interplay between ferromagnetism and unconventional triplet superconductivity has been
mainly studied on the uranium heavy fermion compounds, UGe$_2$~\cite{Sax00}, URhGe~\cite{Aok01} and UCoGe~\cite{Huy07}.
The response to pressure and field
which tune the ground state between superconductivity and the ferromagnetic singularity
is different for all systems.
To date, it is only in UGe$_2$ that the Fermi surface has been experimentally determined
since the large high quality single crystal can be easily obtained for de Haas-van Alphen experiments.~\cite{Sat92b}
Detection of quantum oscillations for the other systems will open up new perspectives.

In particular, the weak itinerant ferromagnet UCoGe with TiNiSi-type orthorhombic structure
is an interesting case.
The ordered moment directed along the $c$-axis is small, $M_0 \approx 0.05\,\mu_{\rm B}$, 
and the Curie temperature $T_{\rm Curie}$ is $\approx 2.6\,{\rm K}$,~\cite{Huy07}
far lower than the characteristic Fermi temperature $T_{\rm F}$
which can be associated with the renormalized band mass.
Superconductivity appears below $T_{\rm sc}\approx 0.6\,{\rm K}$.
Applying pressure leads to a transition from the ferromagnetic state to the paramagnetic state 
at $P_{\rm c}\approx 1.2\,{\rm GPa}$ with the peculiarity,
compared with the two other cases on UGe$_2$ and URhGe,
that superconductivity persists in the paramagnetic phase.~\cite{Has08_UCoGe,Slo09,Has10}
The absence of a deep minimum of $T_{\rm sc}$ at $P_{\rm c}$
seems to reflect the fact that 
for ferromagnetic systems there is not a second order quantum critical point at least in zero field 
but a first order transition associated with a volume discontinuity.
The small size of $M_0$ leads to a small sublattice magnetization
and volume discontinuities and thus to the possibility of preserving superconductivity
in the paramagnetic phase.
Furthermore, a recent approach with a group symmetry theory also indicates that
the transition from a ferromagnetic superconducting state to a paramagnetic superconducting state
should be of first order near $P_{\rm c}$.~\cite{Min08}
A striking additional point is the detection of a very huge and highly anisotropic 
superconducting upper critical field $H_{\rm c2}$
when the field is applied along the hard-magnetization $b$ and $a$-axis
compared with the easy-magnetization $c$-axis ($H_{\rm c2}^c \ll H_{\rm c2}^{a,b}$).~\cite{Aok09_UCoGe,Huy08}
This suggests that for $H\parallel a$ and $b$-axis the magnetic field induces an enhancement of effective mass
as observed for the field re-entrant superconductivity in URhGe with the identical crystal structure.~\cite{Lev07,Miy08}
A recent theory based on field-tuned Ising spin fluctuations~\cite{Tad_ICHE}, 
in which the density of states is proportional to $\sqrt{H}$~\cite{Ish_ICHE},
explains the huge anisotropy of $H_{\rm c2}$ in UCoGe with $H_{\rm c2}^c \ll H_{\rm c2}^{a}$
for the point node gap.
The absence of paramagnetic limitation for $H\perp c$ has been recently explained by 
the itinerant ferromagnetic band splitting.~\cite{Min10}

In this letter, we report the first quantum oscillation study and high field studies of UCoGe
using high quality single crystals.
While the detected Fermi surface is small in volume, 
the cyclotron mass is exceptionally large
implying that UCoGe is a low carrier system with heavy quasi-particles.
This resembles the case of the well-known heavy fermion superconductor URu$_2$Si$_2$.

High quality single crystals of UCoGe were grown using the Czochralski pulling method in a tetra-arc furnace.
The details are described elsewhere.~\cite{Aok_ICHE}
The resistivity shows a clear kink at $2.8\,{\rm K}$ due to the ferromagnetic transition
and the residual resistivity ratio between room temperature and $0\,{\rm K}$ 
($\mbox{RRR}\equiv \rho_{\rm RT}/\rho_0$) was 30
indicating the high quality of the present sample.
The resistivity measurements under high magnetic fields were performed using two experimental setups.
The temperature sweeps down to $80\,{\rm mK}$ at constant fields up to $16\,{\rm T}$ were done using a conventional superconducting magnet.
The field sweeps up to $34\,{\rm T}$ at temperatures down to $40\,{\rm mK}$ were performed 
using a resistive magnet at the LNCMI-Grenoble.
For both experimental setups, the resistivity was measured by a four probe AC method on the same sample.
The field direction was carefully controlled by the single axis-rotation mechanism in both experimental setups.
The electrical current was applied along the $c$-axis and the field direction was rotated from the $b$ to the $c$-axis.

\begin{figure}[tbh]
\begin{center}
\includegraphics[width=1 \hsize,clip]{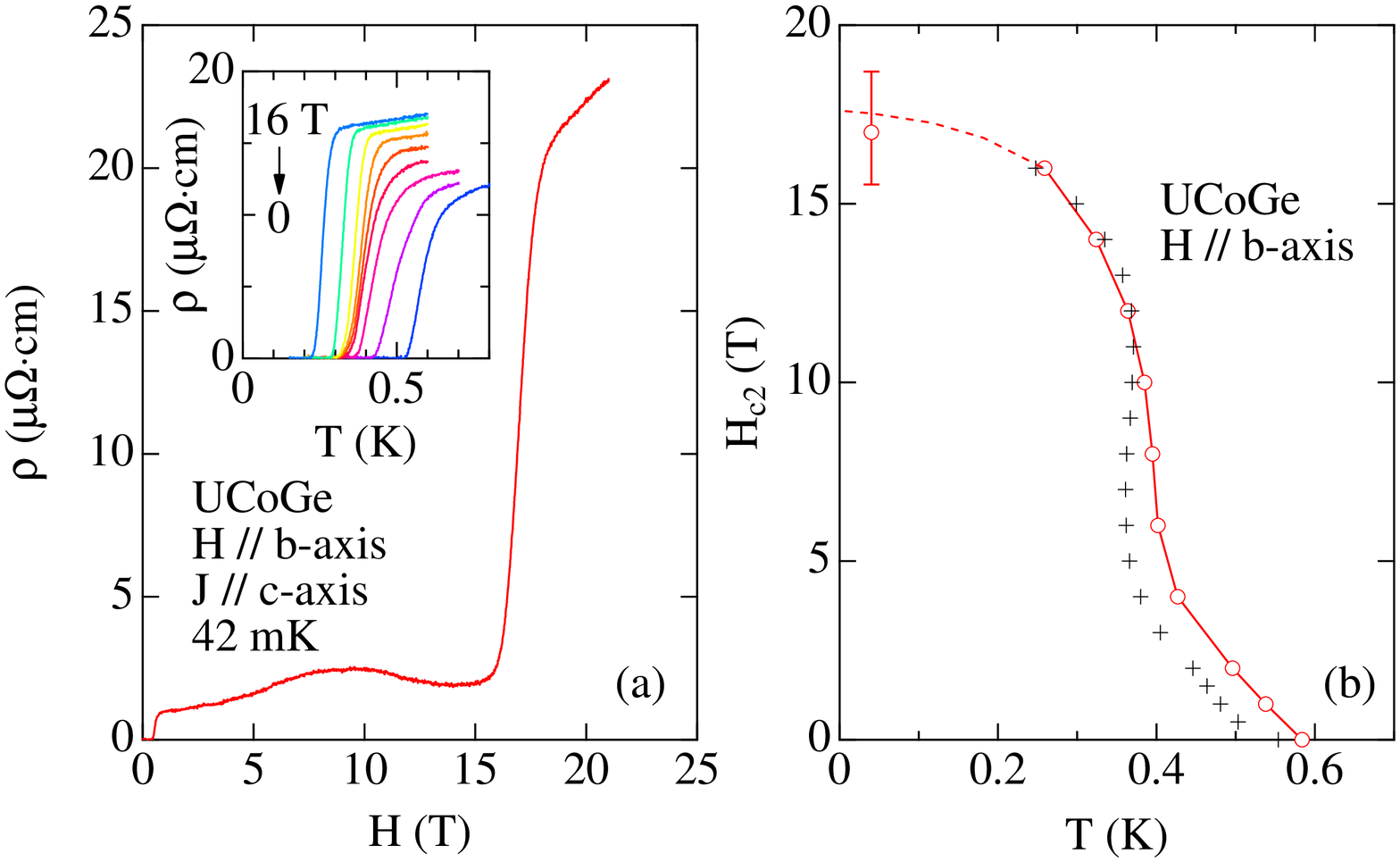}
\end{center}
\caption{(Color online) (a)Field dependence of the magnetoresistance at $42\,{\rm mK}$ for $H \parallel b$-axis in UCoGe. The inset shows the temperature dependence of the resistivity at constant fields from $16$ to $0\,{\rm T}$ in $2\,{\rm T}$ steps. (b)Temperature dependence of the upper critical field $H_{\rm c2}$. The red circles are results of the present study, the black crosses are cited from our previous report in a different sample.~\protect\cite{Aok09_UCoGe} The lines are a guide to the eye.}
\label{fig:UCoGe_b_Hc2}
\end{figure}

The main panel of Fig.~\ref{fig:UCoGe_b_Hc2}(a) shows the high-field magnetoresistance 
for $H \parallel b$-axis at $42\,{\rm mK}$ with an excitation current of $200\,\mu{\rm A}$.
The large jump around $17\,{\rm T}$ indicates the transition from the superconducting state to the normal state,
and corresponds to $H_{\rm c2}$.
The finite value of resistivity below $17\,{\rm T}$, even in the superconducting state,
is most likely attributed to the flux-flow resistivity.
This is caused by energy dissipation due to the vortex moving
without strong pinning centers.
The flux-flow resistivity is also observed in UGe$_2$~\cite{Hux01}.
In fact, in the temperature sweeps of the resistivity with a smaller current of $50\,\mu{\rm A}$,
we detected zero resistivity in the same sample even at $16\,{\rm T}$, as shown in the inset of Fig.~\ref{fig:UCoGe_b_Hc2}(a).
Another possible reason for the finite resistivity below $17\,{\rm T}$ is the inclusion
of a small mis-oriented crystal in the measured sample and the small mis-orientation for the field direction.

Figure~\ref{fig:UCoGe_b_Hc2}(b) shows the temperature dependence of the upper critical field $H_{\rm c2}$ defined 
as the mid point of the resistivity drop.
The $H_{\rm c2}$ curve is in good agreement with the previous results~\cite{Aok09_UCoGe}
although $T_{\rm sc}$ is slightly shifted and the $S$-shaped behavior becomes broader.
In the previous report, we could not clearly indicate $H_{\rm c2}(0)$, 
which exceeded our maximum field of $16\,{\rm T}$ of our superconducting magnet.
In the present study, $H_{\rm c2}(0)$ is clearly demonstrated to be near $17\,{\rm T}$.

\begin{figure}[tbh]
\begin{center}
\includegraphics[width=0.7 \hsize,clip]{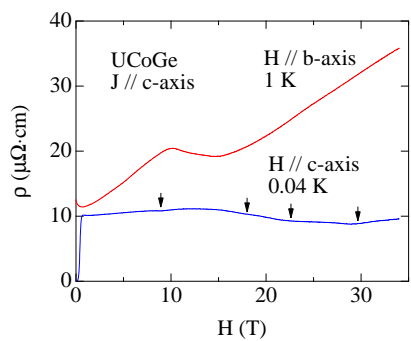}
\end{center}
\caption{(Color online) Field dependences of the resistivity of UCoGe for $H \parallel b$-axis and $c$-axis at $1\,{\rm K}$ and $0.04\,{\rm K}$, respectively.}
\label{fig:UCoGe_MR}
\end{figure}

Let us now focus on the results of the normal state properties.
Figure~\ref{fig:UCoGe_MR} shows the field dependence of the magnetoresistance 
for $H\parallel b$-axis and $c$-axis
which corresponds to the transverse magnetoresistance and longitudinal magnetoresistance respectively.
The magnetoresistance for $H\parallel b$-axis at $1\,{\rm K}$ increases with increasing field,
with a maximum around $10\,{\rm T}$.
This anomaly most likely corresponds to the anomalies detected in the previous experiment,~\cite{Aok09_UCoGe}
where the magnetoresistance for $H\parallel b$-axis with $J\parallel a$-axis 
shows broad kinks between $11$ and $14\,{\rm T}$ with different temperatures.
In the field-temperature phase diagram
these high-field anomalies converge on $T_{\rm Curie}$ at low fields.
In other words, $T_{\rm Curie}$ gradually decreases with fields and disappears around $14\,{\rm T}$
for $H\parallel b$-axis.
A recent theory based on the Landau free energy analysis 
explains that $T_{\rm Curie}$ decreases
when the field is applied perfectly perpendicular to the magnetic easy-axis.~\cite{Min10_arxiv}

In URhGe, it is observed that $T_{\rm Curie}$ for $H\parallel b$-axis decreases with field~\cite{Miy08}
and is connected to $H_{\rm R}\approx 12\,{\rm T}$ at low temperature
where the tilted magnetic moment re-orients to the $b$-axis direction for $H > H_{\rm R}$.
In UCoGe, there is no experimental evidence that the present anomalies is also linked to a spin re-orientation.
However, the key point is that
the magnetic field for $H\perp c$-axis drives the system towards a ferromagnetic singularity.

Contrary to the result for $H\parallel b$-axis, 
the longitudinal magnetoresistance for $H\parallel c$-axis is almost constant with field
although several small anomalies indicated by arrows in Fig.~\ref{fig:UCoGe_MR} are observed.
One can speculate two possible origins for these anomalies.
Firstly, it may be linked to the field-response of the magnetism.
Recent polarized neutron diffraction for $H\parallel c$-axis reveals that
the antiparallel moment ($\sim 0.2\,\mu_{\rm B}$) is induced on the Co site at $12\,{\rm T}$.~\cite{Pro10}
Thus, a complex magnetic field response might affect the magnetoresistance.
The second possible origin is a so-called Lifshitz transition which is related to the topological change of Fermi surface under high magnetic fields related to the band splitting due to the Zeeman effect. 
A similar Fermi surface reconstruction is also reported in the low-carrier heavy fermion superconductor URu$_2$Si$_2$ by Hall resistivity~\cite{Shi09}, thermoelectric power~\cite{Lia_URu2Si2},
and Shubnikov de-Haas (SdH) oscillation.~\cite{Shi09,Has10_URu2Si2}

\begin{figure}[tbh]
\begin{center}
\includegraphics[width=0.7 \hsize,clip]{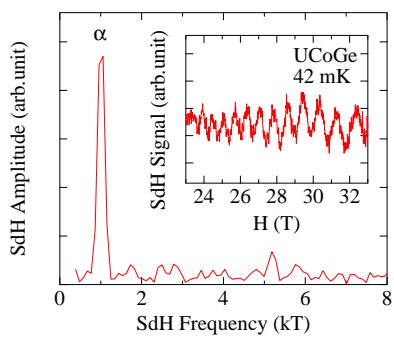}
\end{center}
\caption{(Color online) Typical FFT spectrum for the field tilted 10 deg from $b$ to $c$-axis in UCoGe. The inset shows the corresponding SdH oscillation.}
\label{fig:UCoGe_osc_FFT}
\end{figure}

\begin{figure}[tbh]
\begin{center}
\includegraphics[width=0.7 \hsize,clip]{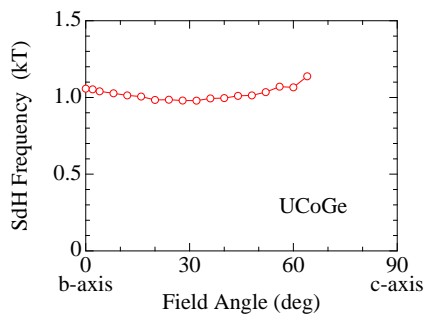}
\end{center}
\caption{(Color online) Angular dependence of the SdH frequency for the field along $b$ to $c$-axis in UCoGe.}
\label{fig:UCoGe_AngDep}
\end{figure}

Now we present the results of the quantum oscillations, 
namely the SdH oscillations at high fields above $20\,{\rm T}$.
Figure~\ref{fig:UCoGe_osc_FFT} represents a typical SdH oscillation and 
the corresponding fast Fourier transformation (FFT) spectrum. 
The background of the magnetoresistance was subtracted using a polynomial function
in order to obtain the oscillatory signal.
The FFT spectrum reveals a single SdH branch $\alpha$ at $1\,{\rm kT}$.
Changing the field angle from $H\parallel b$ to $c$-axis,
the frequency shows no significant change with a slight minimum around $30^\circ$,
as shown in Fig.~\ref{fig:UCoGe_AngDep}
which indicates a small pocket Fermi surface.
It must be noted that the change in the configuration of the magnetoresistance
from transverse to longitudinal with increasing field angle
leads to a strong decrease of the magnetoresistance,
as shown for $H\parallel b$ and $c$-axis in Fig.~\ref{fig:UCoGe_MR}.
The suppression of magnetoresistance is unfavorable for detecting the SdH signal.
Thus, the SdH amplitude is also gradually suppressed with field angle,
and finally the signal becomes undetectable above $65\,{\rm deg}$.

From the temperature dependence of the SdH amplitude for the field tilted $10\,{\rm deg}$ from the $b$ to $c$-axis, for the field range from $22$ to $34\,{\rm T}$,
we determined the cyclotron effective mass $m_{\rm c}^\ast$ at the effective field of $27\,{\rm T}$.
The obtained cyclotron mass for branch $\alpha$ is very large, $25\,m_0$,
although the SdH frequency is relatively small.
Assuming a spherical Fermi surface, the Sommerfeld coefficient $\gamma$ is described as
$\gamma = (k_{\rm B}{}^2 V/3\hbar^2) m_{\rm c}^\ast k_{\rm F}$, where 
$V$ is molar volume, $k_{\rm F}$ is Fermi wavenumber obtained from the SdH frequency
$F=(\hbar c/2\pi e)\pi k_{\rm F}{}^2$.
The $\gamma$-value from specific heat experiments, $\gamma_{\rm Cp}$ is a sum
of $\gamma$-values contributed from each Fermi surface.~\cite{Aok00c}
The cyclotron mass $25\,m_0$ and the SdH frequency $1\,{\rm kT}$ yield
$\gamma = 7\,{\rm mJ/K^2 mol}$.
This value is still not enough to account for the total $\gamma$-value,  $\gamma_{\rm Cp}=55\,{\rm mJ/K^2 mol}$ at zero field obtained by specific heat experiments,
indicating that there are still undetected Fermi surfaces with large cyclotron masses.
It is worth noting that $\gamma_{\rm Cp}$ at $27\,{\rm T}$ should be reduced,
due to the field dependence of the $A$ coefficient of resistivity, as discussed below.
If we assume $\gamma_{\rm Cp}$ at $27\,{\rm T}$ is half of the value at zero field,
the detected value $\gamma = 7\,{\rm mJ/K^2 mol}$ at $27\,{\rm T}$ from one spin-splitting band
is not so low by comparison to the total value.

The Fermi surface corresponding to branch $\alpha$ occupies only $2\,\%$ of the volume of the Brillouin zone,
assuming a spherical Fermi surface.
In general, a large Fermi surface has a large cyclotron mass.~\cite{Onu88}
The value of $2\,\%$ is unusually small, although the cyclotron mass is very large
implying that UCoGe is a low carrier system with heavy effective masses.
This situation also resembles the case of the well known heavy fermion superconductor URu$_2$Si$_2$,
where the multiple small pockets Fermi surface possesses a heavy mass, for example, 
branch $\beta$ ($0.4$--$1.6\,\%$ of the volume of the Brillouin zone) has a mass of approximately $23\,m_0$.

UCoGe is a compensated metal with equal carrier numbers of electrons and holes
since four molecules exist in the unit cell.
A recent band calculation based on the 5\textit{f} itinerant model
predicts small pocket Fermi surfaces, indicating it is a semimetal in the paramagnetic state.~\cite{Sam10}
In the ferromagnetic state, the calculated Fermi surfaces with the spin and orbital polarization
are more metallic
but the carrier number is still small.
For $H \parallel b$-axis in the ferromagnetic state,
two dHvA (SdH) frequencies which come from a closed Fermi surface 
are predicted, $1.7$ and $1.4\,{\rm kT}$.
These values are close to the present results.

Furthermore, the thermopower experiment~\cite{Lia10} reveals a large value of 
$S/T$ in zero temperature limit.
This is consistent with the present result in terms of low carrier with heavy mass,
since the low temperature Seebeck coefficient corresponds to the density of states per carrier.
If the carrier number is small, 
the value of $(S/T)/\gamma$
is large, as reported in URu$_2$Si$_2$~\cite{Zhu09}.

\begin{figure}[tbh]
\begin{center}
\includegraphics[width=0.7 \hsize,clip]{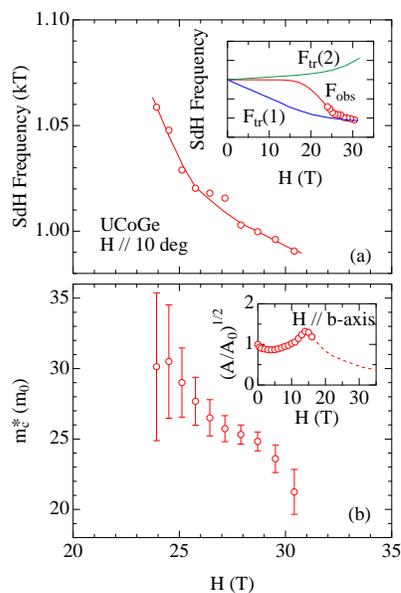}
\end{center}
\caption{(Color online) (a)Field dependence of the observed SdH frequency for the field tilted $10\,{\rm deg}$ from $b$ to $c$-axis in UCoGe. 
The inset of panel (a) shows the schematic field dependence of true frequencies 
$F_{\rm tr}(1)$ for the case 1 and $F_{\rm tr}(2)$ for the case 2.
The line denoted by $F_{\rm obs}$ is expected both from $F_{\rm tr}(1)$ and from $F_{\rm tr}(2)$.
(b)Field dependence of the cyclotron effective mass. Error bars are from the fitting of the mass plot.
The inset of panel (b) shows the field dependence of square root of resistivity $A$ coefficient normalized by zero field value $A_0$ for $H\parallel b$-axis, cited from ref.\protect\citen{Aok09_UCoGe}. A dashed line, which is qualitatively speculated from URhGe, is a guide to the eye.}
\label{fig:UCoGe_Hdep_F_mass}
\end{figure}

For further analysis, we have checked the field dependence of the SdH frequency.
As shown in Fig.~\ref{fig:UCoGe_Hdep_F_mass}(a), the observed SdH frequency strongly decreases with fields
($\Delta F_{\rm obs}/F_{\rm obs} \approx 7\,\%$ from $24$ to $30\,{\rm T}$).
It must be noted that the observed frequency $F_{\rm obs}$ does not directly correspond to
the true frequency $F_{\rm tr}$ when $F_{\rm obs}$ is field-dependent,
because $F_{\rm obs}$ is the back projection of $F_{\rm tr}$ to zero field.
It is described by $F_{\rm obs}=F_{\rm tr} - H\, dF_{\rm tr}/dH$~\cite{Rui82}.
Here we use $H$ instead of $B$, since the ordered moment is small.
Two cases are considered as speculation of the field dependent $F_{\rm tr}$.
The first case is that $F_{\tr}$ decreases with field as shown in the inset of Fig.~\ref{fig:UCoGe_Hdep_F_mass}(a)
denoted as $F_{\rm tr}(1)$.
The volume of the Fermi surface shrinks as a function of the field.
The second case is that $F_{\tr}$ increases with field, which is shown as $F_{\tr}(2)$.
Since there are no reports on high field magnetization,
we cannot definitely choose the one which is the correct case. 
However, if we assume the magnetization curve of UCoGe is similar to that of URhGe,
that is the magnetization for $H\parallel b$-axis may be saturated above $\approx 10$--$14\,{\rm T}$,
the former case $F_{\rm tr}(1)$ is plausible.
Namely, the detected Fermi surface, 
which might be attributed to the minority Fermi surface due to the Zeeman spin splitting effect,
shrinks with field,
while another undetected Fermi surface from majority Fermi surface expands.
It is worth noting that similar field-dependent frequencies are reported in several heavy fermion compounds,
such as CeRu$_2$Si$_2$~\cite{Taka96}, UPt$_3$~\cite{Kim00} and YbRh$_2$Si$_2$~\cite{Rou08}

The cyclotron effective mass also shows field dependent behavior
as shown in Fig.~\ref{fig:UCoGe_Hdep_F_mass}(b).
With increasing field, the cyclotron mass decreases from $30\,m_0$ at $24\,{\rm T}$
to $21\,m_0$ at $30\,{\rm T}$.
This is consistent with the field dependent $A$ coefficient of resistivity, 
as shown in the inset of Fig.~\ref{fig:UCoGe_Hdep_F_mass}(b).
From the Kadowaki-Woods ratio, 
which is roughly valid when the local magnetic fluctuations play a main role,
the $\gamma$-value or the cyclotron mass $m_{\rm c}^\ast$
can be evaluated as $m_{\rm c}^\ast \propto \gamma \propto \sqrt{A}$.
Above $14\,{\rm T}$, $\sqrt{A}$ slightly decreases and it should be expected to continuously decrease
because the superconductivity governed by the orbital limit based on the spin-triplet state 
is destroyed at fields above $17\,{\rm T}$.
The decrease of $m_{\rm c}^\ast$ corresponds to a continuous decrease of $A$.

In summary, high quality single crystals of the ferromagnetic superconductor UCoGe 
were successfully grown
and the high field properties were studied by magnetoresistance measurements.
The huge $H_{\rm c2}$ at low temperature was confirmed for $H \parallel b$-axis.
We observed the first quantum oscillations of UCoGe, which reveal
the small pocket Fermi surface with large cyclotron mass
implying a low carrier system with a heavy electronic state is realized in UCoGe.
The observed frequency decreases with increasing fields meaning the possible shrinkage of the volume of the Fermi surface
with a non-linear field response
and the possibility of a decoupling between minority and majority spin Fermi surfaces.
Correspondingly, we detected a decrease of the cyclotron mass which is consistent with the field-dependent $A$ coefficient and the huge upper critical field $H_{\rm c2}$.

We thank H. Harima, J. P. Brison, L. Howald, L. Malone, W. Knafo, F. Hardy, K. Ishida, Y. Tada and S. Fujimoto for helpful discussion.
This work was supported by the EC program ``Transnational Access'' (EuromagNET II),
ERC starting grant (NewHeavyFermion) and French ANR project (CORMAT, SINUS).


\end{document}